\documentclass[preprint2]{aastex}
\usepackage{graphicx}
\usepackage{color}
\usepackage{amsmath}

\title{Principal Component Analysis studies of turbulence in optically thick gas}

\author{C. Correia$^{1}$, A. Lazarian$^{2}$, B. Burkhart$^{3}$, D. Pogosyan$^{4}$, and J. R. De Medeiros$^{1}$}

\affil{$^{1}$~Departamento de F\'isica Te\'orica e Experimental, Universidade Federal do Rio Grande do Norte, 59072-970, Natal, Brazil; e-mail: \url{caioftc@dfte.ufrn.br}}
\affil{$^{2}$~Astronomy Department, University of Wisconsin, Madison, 475 N. Charter St., WI 53711, USA}
\affil{$^{3}$~Harvard-Smithsonian Center for Astrophysics, 60 Garden St, MS-20, Cambridge, MA 02138, USA}
\affil{$^{4}$~Canadian Institute for Theoretical Astrophysics, University of Toronto, Toronto, ON, Canada}

\newcommand{\co}{$^{13}$CO}

\newcommand{\apca}{$\alpha_{_{\mathrm{PCA}}}$}
\newcommand{\dvdl}{($\delta{v},L$)}

\begin{abstract}

In this work we investigate the Principal Component Analysis (PCA) sensitivity to the velocity power spectrum in high opacity regimes of the interstellar medium (ISM). For our analysis we use synthetic Position-Position-Velocity (PPV) cubes of fractional Brownian motion (fBm) and magnetohydrodynamics (MHD) simulations, post processed to include radiative transfer effects from CO. We find that PCA analysis is very different from the tools based on the traditional power spectrum of PPV data cubes.
Our major finding is that PCA is also sensitive to the phase information of PPV cubes and this allows PCA to detect the changes of the underlying velocity and density spectra at high opacities, where the spectral analysis of the maps provides the universal -3 spectrum in accordance with the predictions of Lazarian \& Pogosyan (2004) theory. This makes PCA potentially a valuable tool for studies of turbulence at high opacities provided that the proper gauging of the PCA index is made. The later, however, we found to be not easy, as the PCA results change in an irregular way for data with  high sonic Mach numbers. This is in contrast to synthetic Brownian noise data  used for velocity and density fields that show monotonic PCA behavior. We attribute this difference to the PCA's sensitivity to Fourier phase information.

\end{abstract}
\keywords{ISM: structure --- magnetohydrodynamics (MHD) --- methods: numerical}

\begin{document}
\maketitle

\section{Introduction}\label{intro}

Determining the power spectrum of magnetic turbulence in different ISM environments is essential as turbulence affects many physical processes including  star formation,  cosmic rays propagation \citep[see][]{2002cra..book.....S, 2015ASSL..407..253Y}, magnetic reconnection \citep{2015ASSL..407..311L}, transport of mass and energy in the ISM. It is well accepted that the ISM is a turbulent and dynamic environment. To complicate the picture, a galactic magnetic field permeates the ISM \citep[see][]{2004ARA&A..42..211E}. Due to the turbulent nature of the ISM there is no possibility to obtain purely analytical predictions  about the motions of gas and dust.  Therefore it is essential to develop and improve statistical tools in order to obtain comprehensive results on observations and to guide the development of theory.

There are many tools for studying  MHD turbulence in the context of the observational ISM.  These include the  Probability Density Functions \citep{2015arXiv150503855B}, Power Spectrum \citep{2013ApJ...771..123B}, Column density tracers \citep{2012ApJ...755L..19B}, Velocity Channel Analysis (VCA), Velocity Coordinate Spectrum \citep[VCS -- see][]{2000ApJ...537..720L,2004ApJ...616..943L,2006ApJ...652.1348L, 2015arXiv150603448C} among others that have solid theoretical foundations. In particular \citet[henceforth, LP04]{2004ApJ...616..943L} deals with the statistics of observational data at high opacities. VCA for example uses spectral analysis and predicts an universal scaling of -3 for the integrated emissivity maps. The implication of this result is that it may be difficult to recover the density power spectrum from CO in the optically thick limit. For optically thick CO, the VCA may only be successfully applied in the limit of thin velocity channels which is a limitation.

In the context of observational data such as  Position-Position-Velocity (PPV) maps, an empirical technique of Principal Component Analysis (PCA), was introduced to the study of ISM turbulence by \citet[][HS97 for now on]{1997ApJ...475..173H} and largely used since then to detect turbulence in synthetic and observational data \citep{1999ASPC..168..387H, 2002ApJ...566..276B, 2002ApJ...566..289B, 2002ApJ...567L..41B, 2004ApJ...615L..45H, 2006ApJ...643..956H, 2009A&A...504..883B}. PCA has also been used to study velocity anisotropy \citep{2008ApJ...680..420H}, galaxy nuclei \citep{2009MNRAS.395...64S} and protostellar jets \citep{2015arXiv150507966C} to cite a few science cases. PCA was reported to be sensitive to underlying turbulence even at high opacities \citep{2003ApJ...595..824B, 2011ApJ...740..120R}. This poses the question why the spectrum of integrated emissivity maps does not reflect the underlying turbulence, while the PCA can still get the information at high opacities? PCA uses multiple channels of a dataset or multiple observations of a same object and perform a linear transformation that maximizes the variation of the data, allowing one to identify the structures that most contributes to the data variance plus being able to easily separate redundant data and noise.

Its main output is an exponent empirically relating characteristic pseudo velocities and pseudo lengths obtained from the product of PCA, the eigen images and eigen vectors, $\delta{v}\sim{}L^{\alpha}$. This is one way to relate the variation of kinetic energy with size scale in turbulent flows. Thus the exponent $\alpha$ can also be related to the spectral index $\beta$ of the PPV's integrated intensity of the cloud and to its structure function \citep[see][for a detailed description of the technique]{2002ApJ...566..276B,2002ApJ...566..289B}. This relation was found in observational studies such as HS97; \citet{2002ApJ...566..289B,2004ApJ...604..196B} to cite a few. It seems that $\alpha$ won't be affected by an increase of the ISM opacity and we investigate its response to changes in optical depth. Our claim in this paper is that it occurs because the results from Principal Component Analysis contain the Fourier phase information unlike the spatial power spectrum which only contains amplitude information.

We organize this work as the following: Section \ref{simu} details the synthetic data used in this work; Section \ref{vca} shows the predictions and results of VCA; Sect. \ref{pca} we detail the PCA analysis and its results on different scenarios. In section \ref{disc} we discuss the results and its implications.

\section{Synthetic data}\label{simu}

\begin{figure*} [htp!]
\vspace{-2.0cm}
\centering{\includegraphics[width=0.86\linewidth]{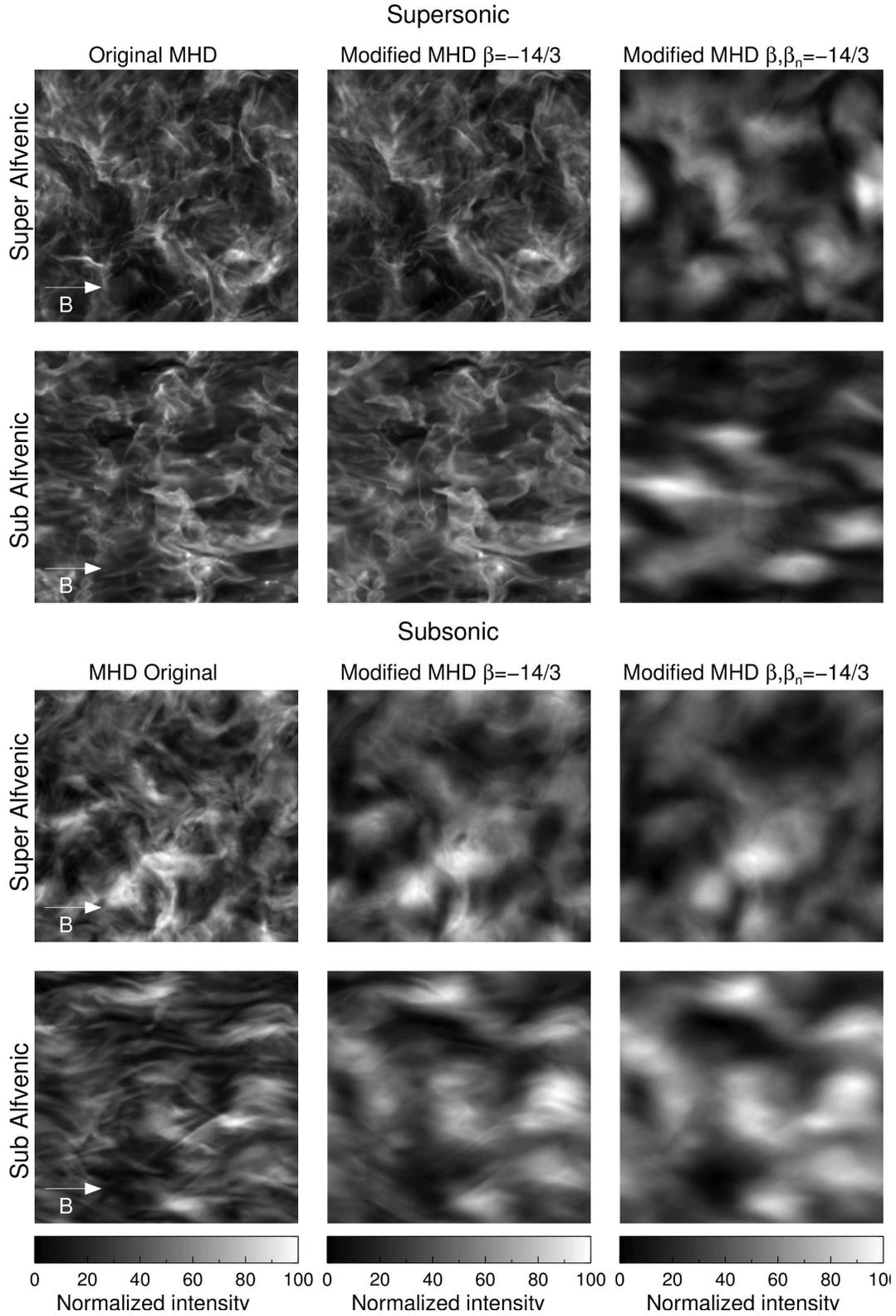}}
%\vspace{-0.2cm}
\figcaption{\footnotesize Integrated intensity maps of the transonic simulations presented in figs. \ref{pwer-mhd}, \ref{pwer-mhdforce} and \ref{pwer-mhdforcedn} in columns 1, 2 and 3 respectively. First column present the original MHD simulations, in the second are the MHD with rescaled velocity spectra and MHD simulations with both density and velocity rescaled are in the third column. First two rows are supersonic simulations and third and fourth are subsonic. Sub-Alfv\'enic and super-Alfv\'enic simulations are shown on the panels on left. White arrow shows magnetic field direction.\label{IImap}}
\end{figure*}

We generate 48 PPV cubes from 12 fractional Brownian motion (fBm) velocity cubes and constant density with $\beta$ ranging from 1.0 to 6.5 and 36 PPV cubes from 12 fBm velocity and density cubes with $\beta=2.0,2.5,3.0,3.5$ and $\beta_n=1.0,1.5,2.0$ with cube sizes of $256^3$; for further inspection on opacity and phases regarding the relation between PCA exponents and the structure function of turbulent clouds found empirically in \citet{2002ApJ...566..276B,2002ApJ...566..289B}.

In addition we use 20 3D numerical simulations with large scale solenoidally driven compressible (MHD) turbulence of the ISM; 12 cube sizes of $256^3$ and 8 of $512^3$; Alfv\'en Mach Number ranging from 0.2 to 15, Sonic Mach Number from 0.7 to 10. The magnetic field consists of a uniform initial field directed towards x-axis and a time-dependent component: $\mathrm{\textbf{B = B}}_{ext}+\mathrm{\textbf{b(t); b(0) = 0}}$. For more details on the simulations scheme see \citet{2002ApJ...564..291C,2003MNRAS.345..325C} and \citet{2009ApJ...693..250B}.

The radiative transfer used in this work is a post-process to the velocity and density cubes with the SimLine3D \citep{2002A&A...391..295O} radiative transfer code to generate synthetic Position-Position-Velocity (PPV) cubes of the \co~$J=2-1$ transition. This code computes the local excitation of molecules from radiation absorption at wavelengths of the molecular transitions and from collisions with the gas using two approximations. First is a computation of a local radiative interaction volume for every point, limited by the velocity gradients, and beyond this volume it computes an interaction of the points with the average radiation field \citep[see][for details on the radiative transfer code]{2002A&A...391..295O,2013ApJ...771..123B}.  Temperature set to $10~K$; cloud size of $5~pc$; density scaling factor ranges of 9, 275, 8250 and 82500 $cm^{-3}$ and a relative abundance $[X/H_2]=1.5E-5$.

We apply a radiative transfer code to 48 fBm velocity cubes with constant density and optical depth $\tau$ set to $\sim$ 0.2; 2; 20 and 200, plus 36 fBm velocity and density cubes with $\tau$ set to $\sim$ 0.2, 2 and 20, resulting in 74 fBm PPV cubes in the total. A radiative transfer is also applied to a set of 12 unmodified MHD simulations of 3D velocity and density cubes, including combinations of subsonic, supersonic and subAlfvenic, superAlfvenic simulations, here mentioned as Original MHD simulations; we also include a set of 36 rescaled velocity but keeping the original density, mentioned as velocity-rescaled MHD simulations; and a set of 24 MHD simulations with both density and velocity rescaled, mentioned as density+velocity-rescaled MHD simulations. Resulting in a total of 72 PPV cubes from MHD simulations. The rescaling of the velocity and density is described in the next subsection.

\subsection{Handling the data}\label{handling}

\begin{figure*} [t!]
\vspace{-1.8cm}
\includegraphics[width=0.95\linewidth]{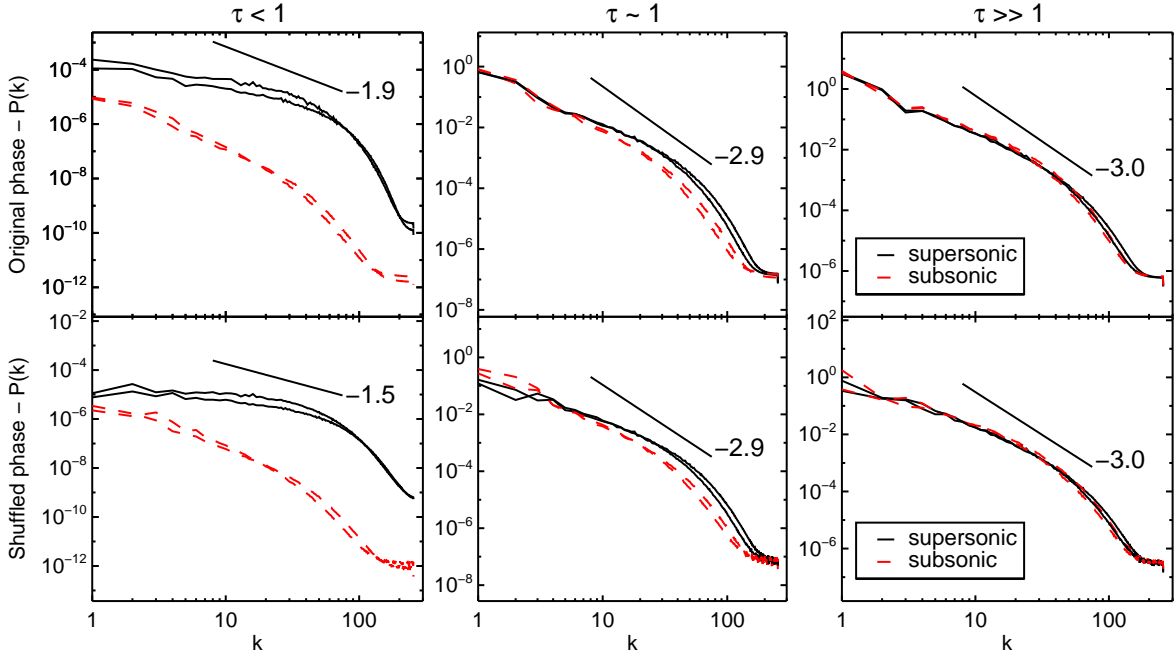}
\vspace{-0.2cm}
\figcaption{\footnotesize PPV total intensity spectrum vs. wavenumber for original MHD simulations. First row shows the simulations with original phases; second row shows the same simulations but with the PPV phases shuffled; Black lines are supersonic simulations; Red lines are subsonic simulations. Straight black lines shows the average slope for the supersonic simulations. \label{pwer-mhd}}
\end{figure*}

In order to better isolate the many variables that can affect the results of our analyses we apply three additional and separated modifications on the data previously described above. 

One is a rescaling of the amplitude $\rho_v$ of Fourier transform of the velocity fields from MHD simulations in order to have one big power law in all its range. It can be obtained by averaging the spectra of all three velocity components -- $\langle{P(k)}\rangle=[P_x(k)+P_y(k)+P_z(k)]/3$ -- and multiplying the amplitudes with the ratio between the desired spectrum and the average spectrum: $\rho'_{vx}=\rho_{vx}\times{P_{\beta}/\langle{P}\rangle}$, where $P_{\beta}(k)\sim{k^{-\beta}}$; $\beta=8/3;~11/3;~14/3$. The process follows in $y$ and $z$ directions. After this modification, a new set of PPVs is generated from radiative transfer. This modification will preserve the differences between the velocity components, e.g., any possible anisotropies.

In another modification we rescale both the density and velocity of MHD simulations to $\beta,\beta_n=8/3;~14/3$. The rescaling is also applied to an additional set of 36 fBm simulations with velocity spectral index $\beta$ varying from 2.0 to 3.5 and density spectral index $\beta_n$ from 1.0 to 2.0.

The last modification is a randomization of the phases of the PPV cubes, obtained by randomly shuffling the phases of the synthetic PPV in the Fourier space and then returning it to the configuration scale in order to probe phase.

\section{Power spectrum results}\label{vca}

In accordance with the prediction of \citet{2004ApJ...616..943L} and the confirmation from \citet{2013ApJ...771..123B} we observe a saturation of the spectral slope at $P(k)\sim{k^{-3}}$ when increasing the opacity of the medium, independently of the strength of the magnetic field or the energy spectrum. Fig. \ref{pwer-mhd} exhibits the power spectrum of the Original MHD simulations for different optical depths in the first row -- subsonic in red and supersonic in black on the online version -- , with the original phases. The optically thin cases (left column) can clearly be distinguished between subsonic (red) and supersonic simulations (black). As we increase the opacity, the difference is very small in transonic simulations (middle) and supersonic cases have indistinguishable spectra (right column). When the phases in Fourier space are randomly shuffled, the overall shape of the spectra won't change (see Fig. \ref{pwer-mhd}, second row) and the behavior described above is repeated. Shuffling the phases of the PPV will produce only small fluctuations on the resulting spectra, this effect is more noticeable in the energy injection range and the optically thin cases.

\begin{figure}[h!]
%\vspace{-1.2cm}
\includegraphics[width=1.0\linewidth]{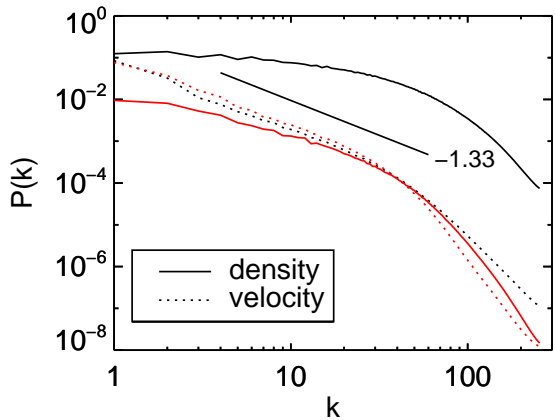}
\figcaption{\footnotesize Energy spectrum vs. wavenumber of the density (solid line) and the average velocity (dotted line) of Original MHD simulations depicted in Fig. \ref{pwer-mhd}. Supersonic simulations are in black and subsonic in red. \label{pwer-mhdcube}}
\end{figure}

Fig. \ref{pwer-mhdcube} shows the energy spectra of the density (solid lines) and velocity (dotted lines) components for supersonic MHD simulations (black) and subsonic MHD simulations (red). It is easy to see that more energy is stored at small scales in the density cubes for the supersonic simulations making the density spectra shallow, while the velocity spectra  changes its slope marginally. This energy concentrated within density is probably due to the concentration of matter by shocks \citep{2005ApJ...624L..93B}. A visual inspection of Fig. \ref{IImap} showing transonic MHD simulations can confirm the above explanation that subsonic simulations are more affected by the rescaling of velocity and that we cannot distinguish previously supersonic and subsonic MHD simulations. 

\begin{figure*}[t!]
\vspace{-1.2cm}
\includegraphics[width=0.96\linewidth]{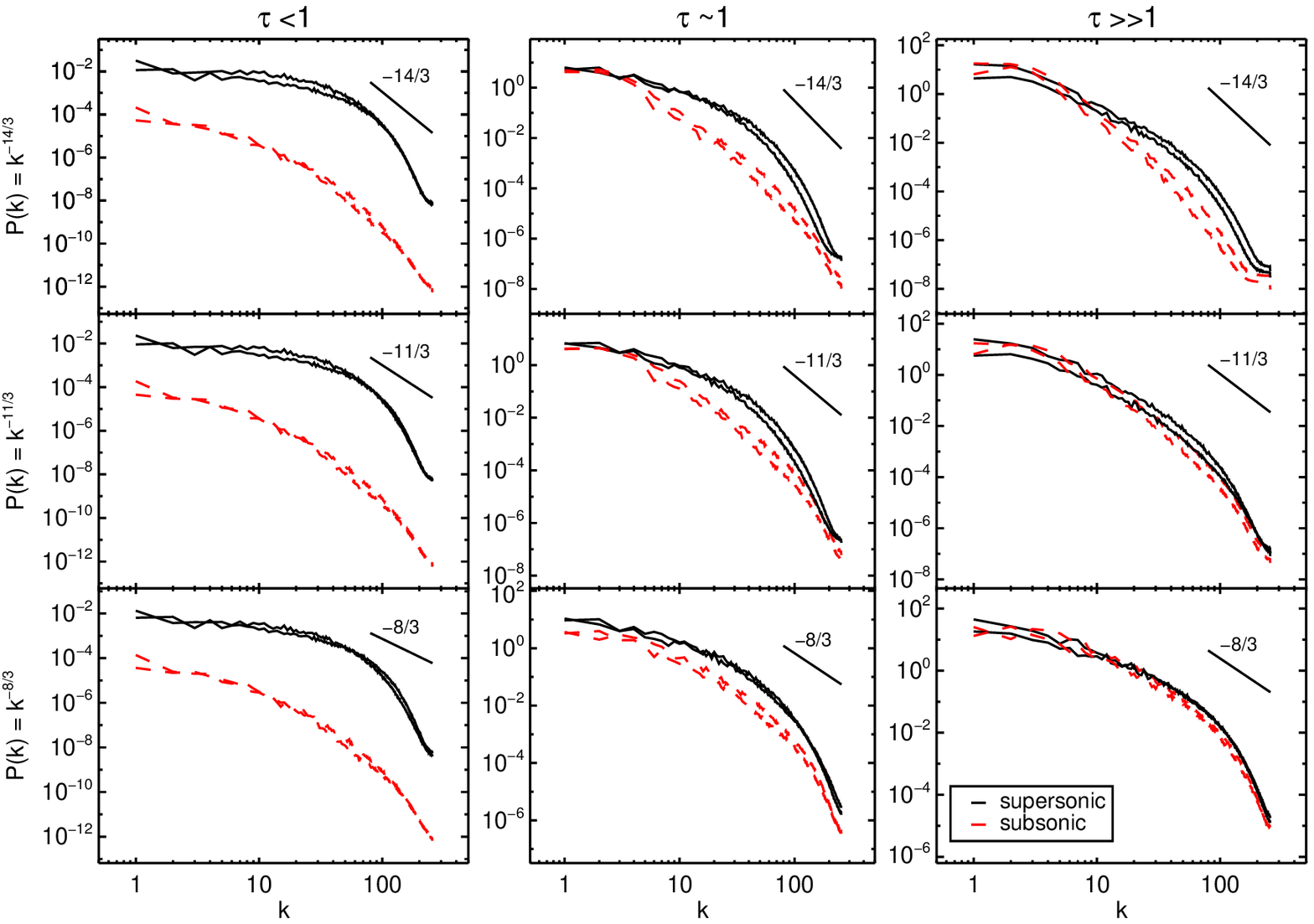}
\vspace{-0.2cm}
\figcaption{\footnotesize PPV total intensity spectrum vs. wavenumber of the velocity-modified MHD simulations. First row is the modification to $P(k)\sim~k^{-14/3}$; Second row to $P(k)\sim~k^{-11/3}$; and third row to $P(k)\sim~k^{-8/3}$;  Black lines are supersonic simulations; subsonic simulations are in red. Straight lines are shown for comparison effects only. \label{pwer-mhdforce}}
\includegraphics[width=0.96\linewidth]{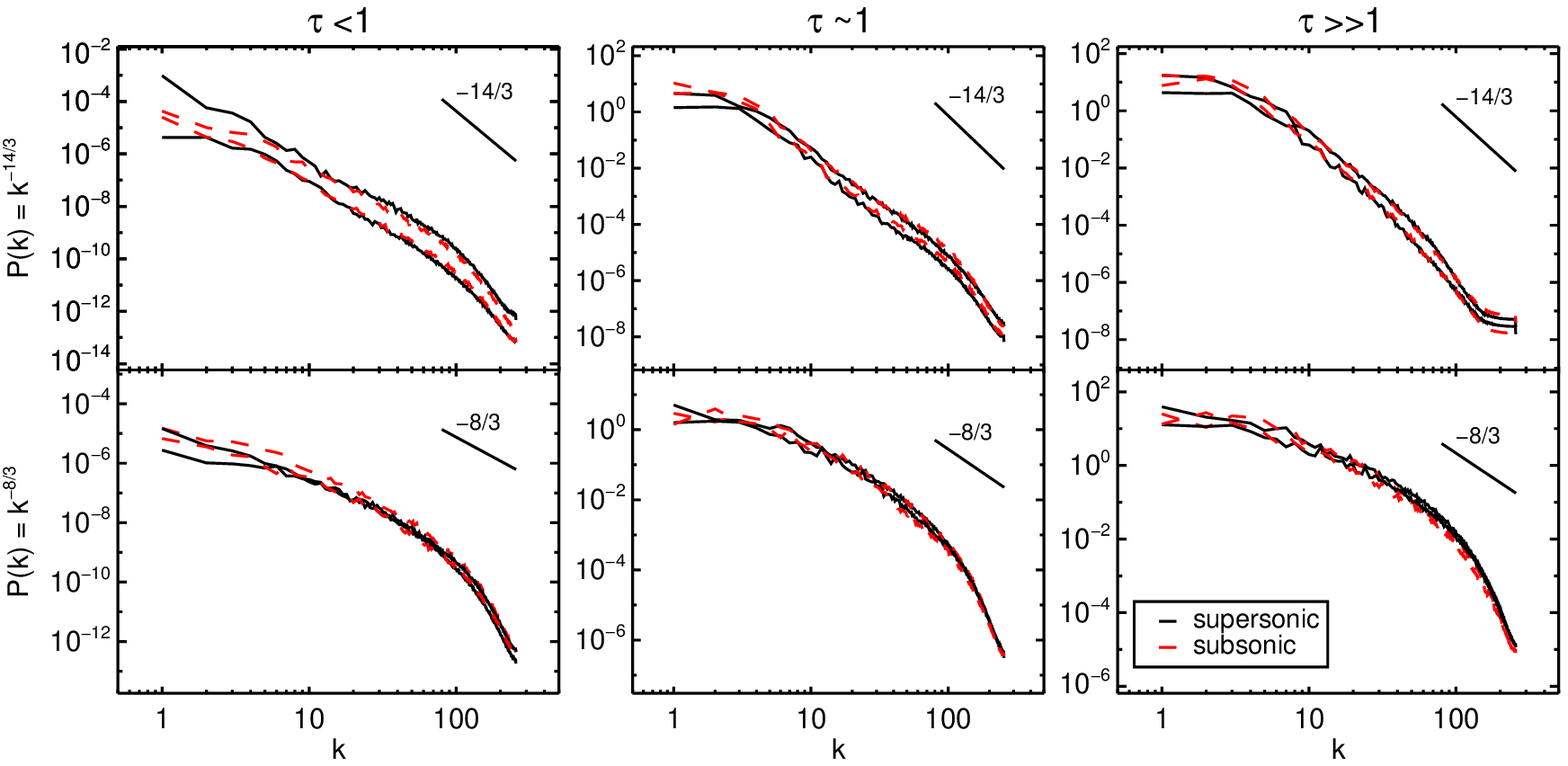}
\vspace{-0.2cm}
\figcaption{\footnotesize PPV total intensity spectrum vs. wavenumber of the density+velocity-modified MHD simulations. First row is $P(k)\sim~k^{-14/3}$; Second row is $P(k)\sim~k^{-8/3}$;  Black lines are supersonic simulations; subsonic simulations are in red. Straight lines are shown for comparison effects only.\label{pwer-mhdforcedn}}
\end{figure*}

Even though we can calculate the energy spectra of the velocity cubes of our MHD simulations within the inertial range, they do not present a perfect power law. To isolate some possible distortion of the resulting spectral analysis or the Principal Component Analysis, we show in Fig. \ref{pwer-mhdforce} the spectra of the velocity-modified MHD simulations described in section \ref{handling}.

The columns are organized from left to right by increasing opacity as in Fig. \ref{pwer-mhd} but each row is represented by a different modification of the velocity power spectra -- from top to bottom: $\beta=14/3;~11/3;~8/3$. It is interesting to note how the subsonic simulations are more affected by the modification of the velocity spectrum, specially on the optically thin cases. Still, it is hard to distinguish the subsonic and supersonic simulations in high opacity cases (right panels).

The same procedure follows to density+velocity-modified MHD simulations -- $\beta,\beta_n=14/3;~8/3$ and it is shown in the figure \ref{pwer-mhdforcedn}. We observe no change in the overall shape of the spectra of PPV derived from velocity-modified MHD simulations or from density+velocity-modified MHD simulations when the phases are randomly shuffled.

One way to separate the contributions of density and velocity to the resulting intensity fluctuations within the PPV cube is the Velocity Channel Analysis (VCA), a technique based on spectral analysis, analytical formulations and numerically tested that cover the spectra of subsonic and supersonic turbulence. By changing the thickness of the velocity channels analyzed on a PPV cube, making it possible to disentangle the effects of density and velocity to the turbulent spectrum. 

\section{PCA study on turbulence}\label{pca}

Applying Principal Component Analysis to a big dataset as a PPV cube reorders the data according to the variance. This empirical technique has been shown to detect turbulence statistics \citep{1999ASPC..168..387H}. In this work we test how PCA behaves at high opacity environments where VCA provides universal spectrum of total intensity fluctuations. We include a short description of this method. Let the PPV cube be denoted as $T_0(r_i,v_k)$ with $n\times{n}$ of size and $n_v$ velocity channels, where $r_i=(x_i,y_i)$ denotes the position on the sky and $v_k$ is the spectral emission on that line of sight. In order obtain only the variance of each velocity channel, we subtract every channel from its average intensity,
\begin{equation}
	T_{ik}=T_0(r_i,v_k)-\frac{1}{n}\sum\limits_{j=1}^{n}{T_0(r_j,v_k)}\label{ppvavg}
\end{equation}
for $i=1,n$ and $k=1,n_v$. Next, we calculate the covariance matrix $S$ of the data
\begin{equation}
	S_{kl}=\frac{1}{n}\sum\limits_{i=1}^{n}{T_{ik}T_{il}}~,\label{covmat}
\end{equation}
One can solve an eigenvalue equation for this covariance matrix: $S\textbf{u}=\lambda\textbf{u}$. This will result in a set of $n_v$ eigenvalues $\lambda$ and eigenvectors $\textbf{u}$ in which the variance of the data is maximized. The eigenvectors will provide the characteristic pseudovelocities. To get information of the spatial variance one must project each eigenvector into the PPV cubes, which will result on a set of $n_v$ eigen images $I_l(r_i)$,
\begin{equation}
	I_l(r_i)=\sum\limits_{k=1}^{n_v}{T_{ik}\textbf{u}_{lk}}~.\label{eigimg}
\end{equation} 
With the eigenimages and aigenvectors in hand we apply an autocorrelation function (ACF) to this data in order to obtain characteristic scales of pseudolenghts and pseudovelocities, since the data transformed via PCA won't represent necessarly these physical quantities, but a combination of such in a space where the variance is maximized. Next, we obtain the autocorrelation functions (ACFs) of the corresponding eigen vectors $\textbf{u}$ and eigen images $I$ of the $l^{th}$ component,
\begin{subequations}
	\begin{eqnarray}
		C^l_V(\delta{v})=\langle{u}^l(v)u^l(v+\delta{v})\rangle~;\\\label{acfv}
		C^l_L({L})=\langle{I}^l(\textbf{r})I^l(\textbf{r}+{L})\rangle~. \label{acfl}
	\end{eqnarray}
\end{subequations}
The characteristic pseudo velocities $\delta{v}$ (scales ${L_B}$) of the $l^{th}$ component are defined as when the ACF of the $l^{th}$ eigen vector (eigen image) drops by one e-fold -- $C^l_V(\delta{v})=1/e$ ($C^l_L(L_B)=1/e$). Due to resolution limitations, the characteristic velocities (scales) are interpolated between the nearest points to $1/e$. The spatial resolution limitation however will result in overestimated values of the \textit{biased} characteristic scales $L_B$ close to the resolution limit, making a correction necessary,
\begin{equation}
L=(L_B^{\kappa}-\epsilon{L_{res}}^{\kappa})^{1/\kappa}
\label{dlcorrection}
\end{equation}
where $L$ is the resolution-corrected characteristic scale; $\epsilon$ depends on the shape of the telescope beam; $L_{res}$ is the resolution limit; and $\kappa$ depends on the shape of the ACF of the eigen images. Velocity doesn't need a resolution correction \citep[see][for an explanation on the resolution corrections]{1999ASPC..168..387H}. Then, we scale the recovered \dvdl~pairs to obtain the scaling exponent $\alpha$
\begin{equation}
\delta{v}=v_0L^{\alpha}~.\label{pcascaling}
\end{equation}
If this correction is not taken into account, one may overestimate the slope ($\alpha$ -- \apca~for now on) in Eq. \ref{pcascaling}. \apca, the PCA scaling exponent can also be interpreted as a pseudo-structure function since it is calculated relating the values of the pseudo \dvdl~pairs. First components are related to higher velocity and scale variances and therefore higher values of \dvdl~pairs \citep{2011ApJ...740..120R}. We recover typically up to 10 or 12 pairs, that will account for almost all variance on the data. Simulations with low values of the energy spectral index $\beta$ present less quantity of \dvdl~pairs that can be recovered above the velocity and length resolution limits. In our analysis, we consider this lower limit of $\beta$ being $\beta<2$ since the number of pairs recovered is usually less than three, resulting in large error of the slope determination and therefore these simulations are excluded from the analysis. Similarly, for $\beta\gtrsim5.0$, the exponent \apca~start to behave erratically and saturate around \apca$\sim1.0$. So we consider our PCA calibration trustworthy within the range $2<\beta<5$.

\begin{figure*}[t!]
\includegraphics[width=1\linewidth]{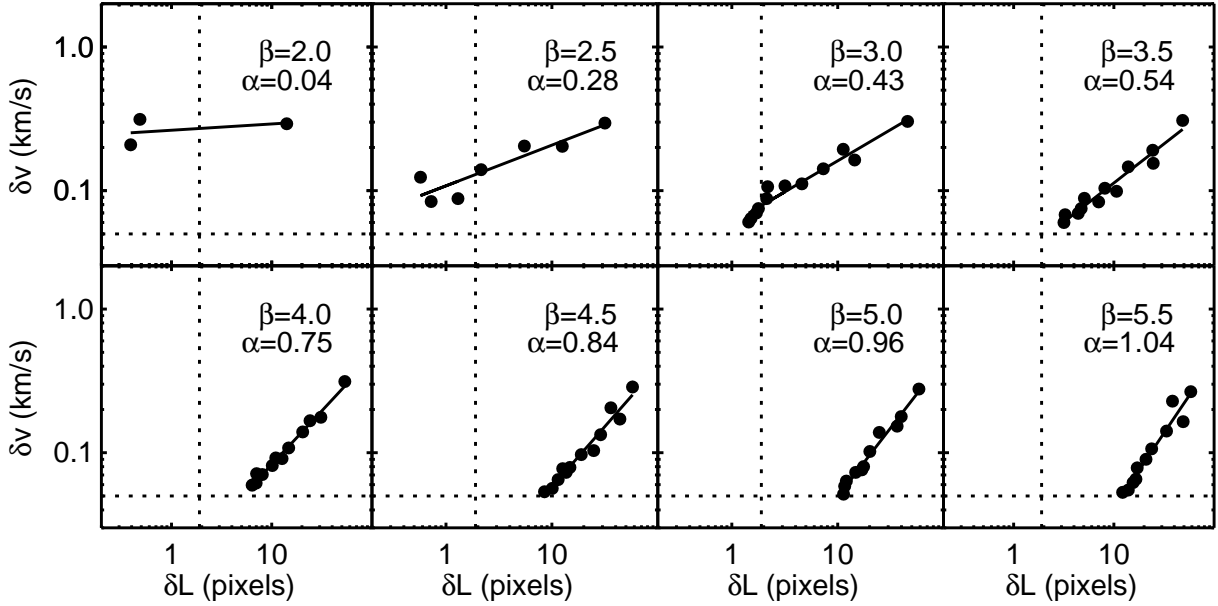}
%\plotone{DVDL-PCA-dn275.ps}
\figcaption{\footnotesize Deriving \apca exponents for fBm simulations (\co, $275~cm^{-3}$, $\tau\sim2$). Each panel shows a different $\beta$ -- 2.0 to 5.5. From equation \ref{pcascaling}, the slope of the linear fit will provide the \apca. dotted lines shows the resolution limits for the velocity (horizontal) and resolution (vertical). \label{dvdl-fbm-cte}}
\end{figure*}

\citet{2013MNRAS.433..117B} provided an analytical calibration of the PCA outputs on a PPV but one must remember that the resulting eigen images and eigen vectors do not represent actual physical features. In a more realistic environment, density and velocity information are tangled within those outputs. Many physical features in the cloud like intermittency, opacity, density fluctuations will affect the resultant emission featured to the observer and therefore the PCA exponents \citep{2011ApJ...740..120R,2014MNRAS.440..465B}.

\subsection{Fractional Brownian motion}\label{fbm}
\begin{figure} [t!]
\includegraphics[width=1\linewidth]{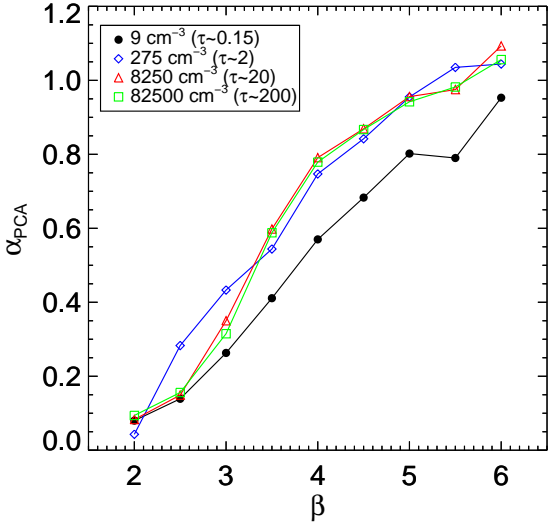}
%\plotone{f3-alphabeta-fbm.ps}
\figcaption{\footnotesize \apca~for fBm simulations as a function of the velocity energy spectrum, $\beta$, for constant density. Different symbols/colors represent different opacities (see legend). \label{abeta-fbm}}
\end{figure}

\placefigure{abeta-fbm}

\begin{figure} [ht!]
\includegraphics[width=1\linewidth]{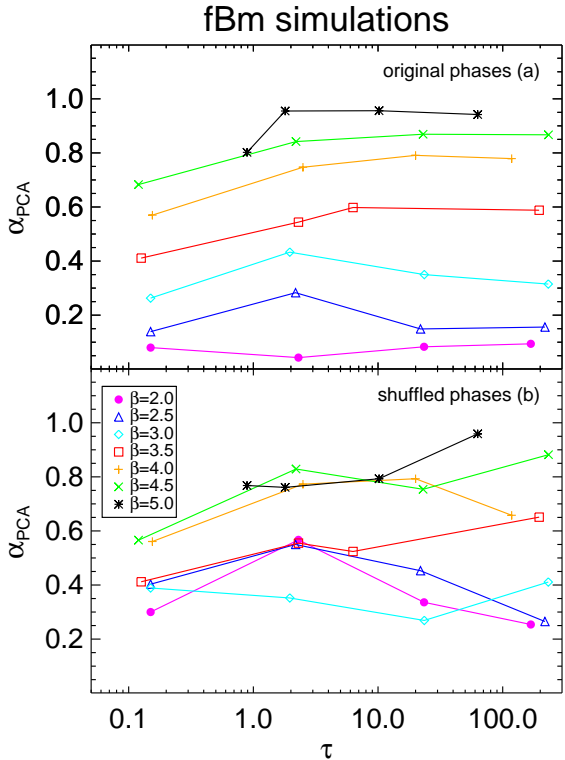}
\figcaption{\footnotesize \apca of fBm clouds with uniform density as a function of the optical depth, $\tau$. In the top panel we kept the original phases. Bottom panel shows the result of shuffling the PPV phases. Different collors represent different spectral indexes. \label{atau-fbm}}
\end{figure}

\placefigure{atau-fbm}

\begin{figure*} [ht!]
\includegraphics[width=1\linewidth]{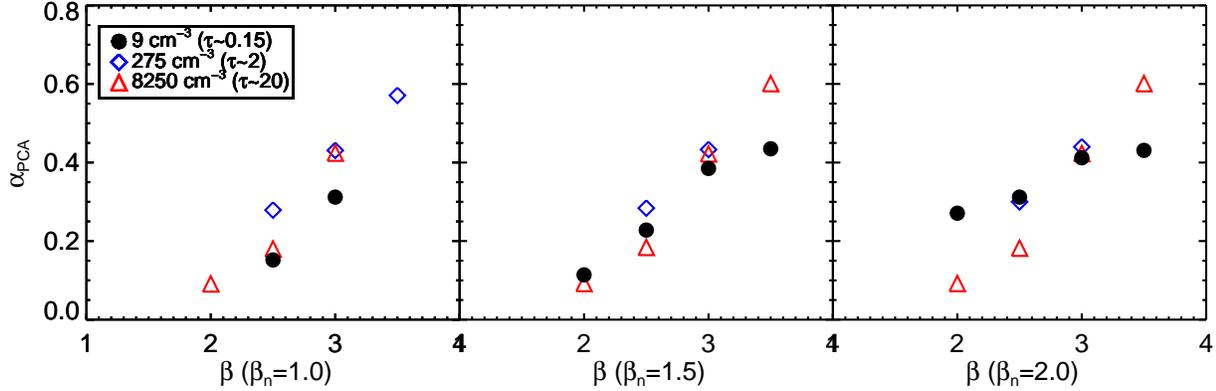}
\figcaption{\footnotesize \apca~for fBm simulations as a function of the velocity energy spectrum, $\beta$, for differente density indexes (from left to right: $\beta=1.0;~1.5;~2.0$). Different symbols/colors represent different opacities (see legend).\label{abeta-fbmdn}}
\end{figure*}

\placefigure{abeta-fbmdn}

Following \citet{2002ApJ...566..276B}, we now apply PCA to synthetic data of fBm velocity cubes and constant density later converted to PPV cubes fitting the characteristic pseudo-velocities $\delta{v}$ to the corresponding component of pseudo-scales $L$ we obtain the exponent $\alpha$ (see eq. \ref{pcascaling} and Fig. \ref{dvdl-fbm-cte}), as described in section \ref{simu}. Note that the resolution correction will push the characteristic scales below the resolution limit. In Fig. \ref{abeta-fbm} we plot \apca~as a function of the energy spectra $\beta$ for fBm simulations. \apca~can be empirically related to the spectral index of the cloud but our goal here is to probe its sensitivity changes in the phase and spectrum.

Plotting \apca~as a function of optical depth $\tau$ in Fig. \ref{atau-fbm}, \apca~will vary very little for a given $\beta$ (see Fig. \ref{atau-fbm}, top panel) for a wide range of opacity (0.2 to 200), suggesting that PCA won't be greatly affected by high opacities. On the other hand, applying a random phase shuffling on the PPV cubes from fBm simulations will partially spoil Principal Component Analysis (\ref{atau-fbm}, bottom panel).

Top panel of Fig. \ref{atau-fbm} exhibit \apca~of PPV cubes from fBm simulations with the original phases, showing that for a given spectral index, \apca~is not greatly affected by the increasing of opacity of the ISM. Bottom panel of Fig. \ref{atau-fbm} shows that shuffling the phase greatly affects PCA results, meaning that PCA also takes advantage of phase information.

The procedure is repeated on PPV from fBm density ($\beta_n$ ranges of 1.0, 1.5 and 2.0) and fBm velocity cubes ($\beta$ ranging from 2.0 to 3.5) and for high opacities, PCA provides the same chaotic behavior but for optically thin simulations, the relationship between \apca~and $\beta$ gets shallower with and increase of $\beta_n$ (See Fig. \ref{abeta-fbmdn}).

\subsection{Magnetohydrodynamics}\label{pca-mhd}
\begin{figure} [t!]
\includegraphics[width=0.85\linewidth]{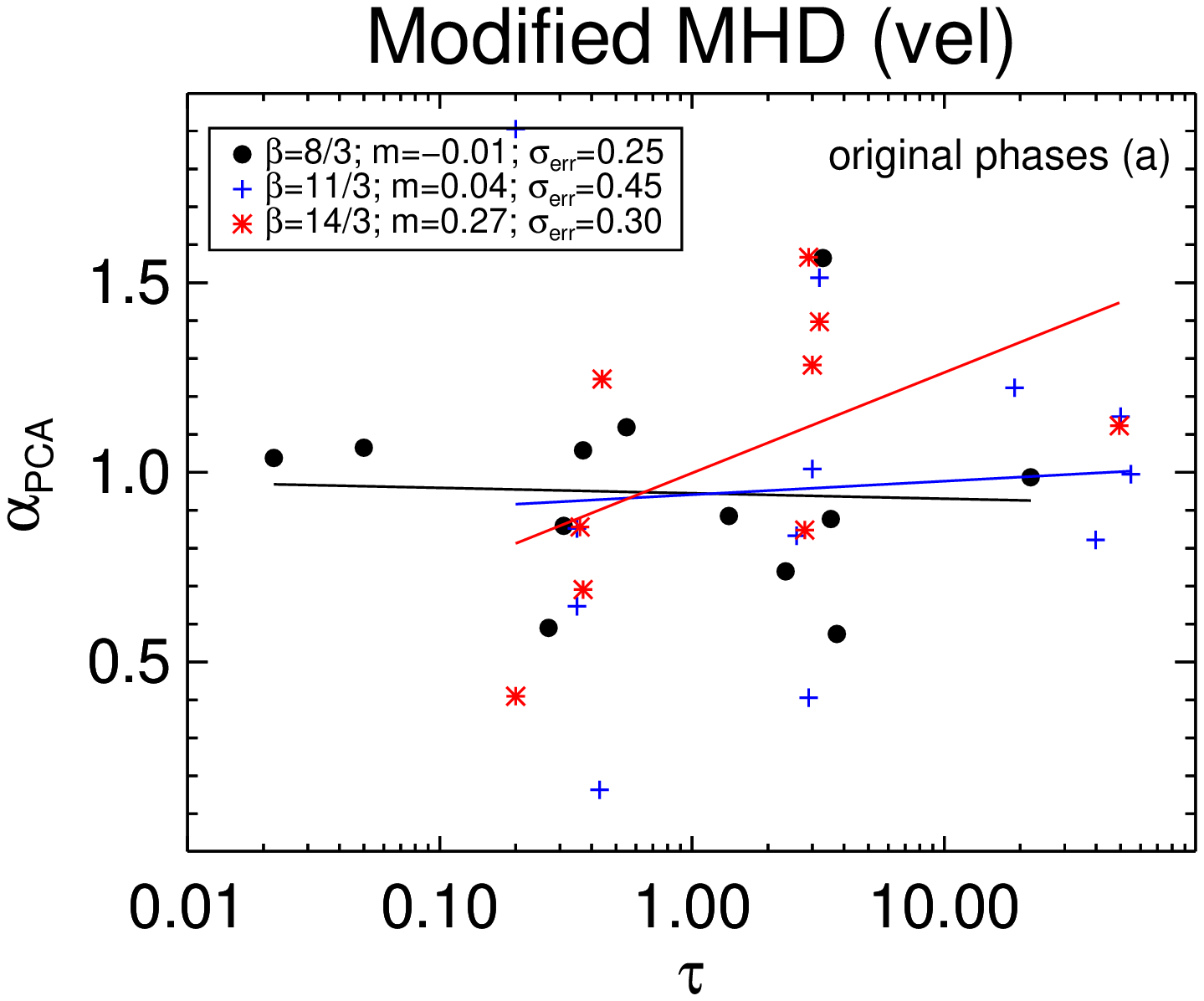}
\includegraphics[width=0.85\linewidth]{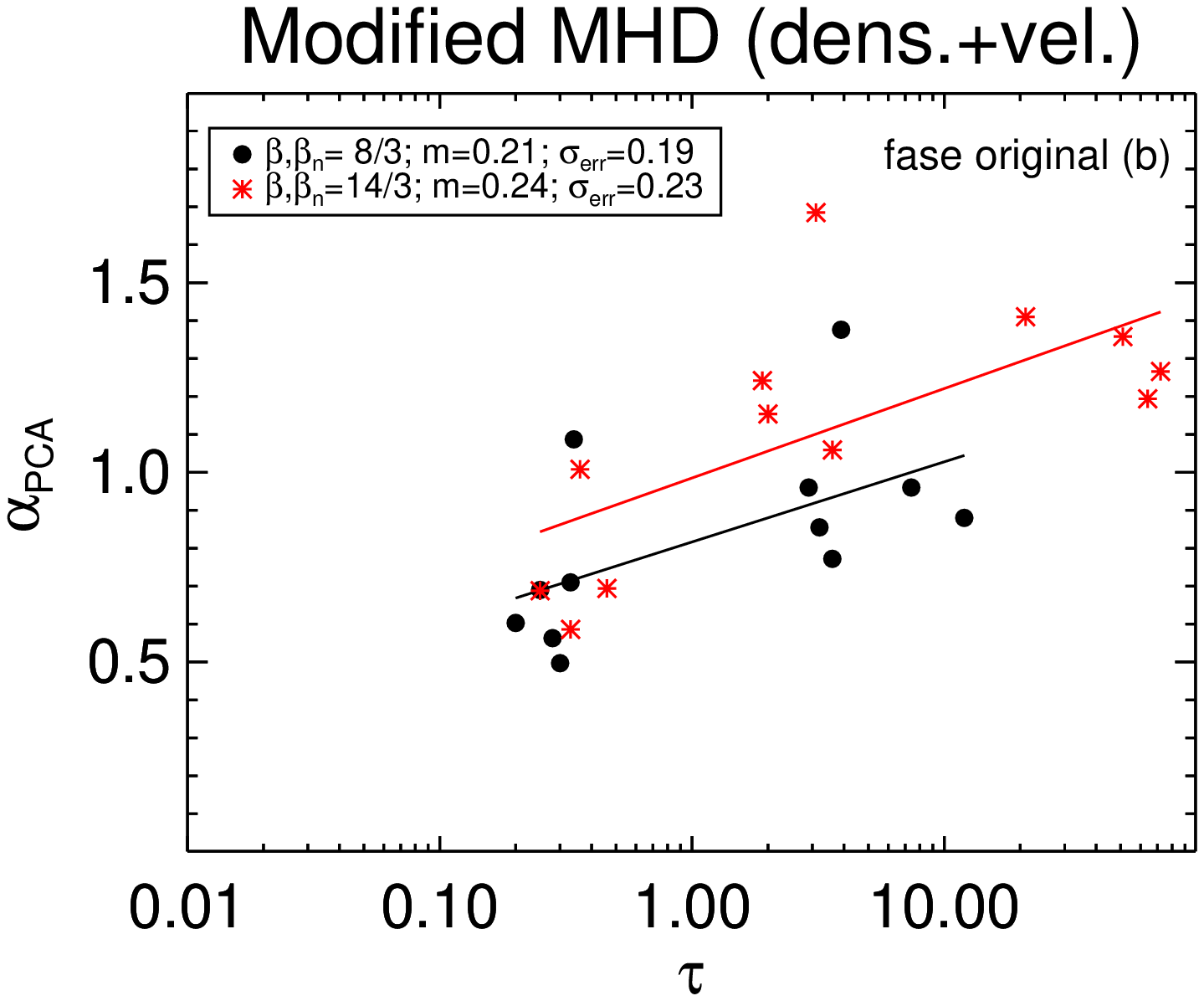}
\figcaption{\footnotesize \apca~$vs.~\tau$ of (a) modified velocity and (b) modified density and velocity on MHD simulations. Only original phases are shown here. Different colors represent different spectral indexes. Lines are fits with minimized $\chi^2$ error. $m$ is the slope and $\sigma_{err}$ the standard error of the regression.\label{atau-MHDforce}}
\end{figure}

When applying PCA to modified MHD simulations were the velocity field is rescaled to a well defined power law we expected that \apca~woudn't be greatly affected by intermittency of the velocity and density fields, specially for the subsonic simulations where the influence of filaments and intermittent are small (see Fig. \ref{IImap}). Instead, when plotting \apca~against optical depth (see figs. \ref{atau-MHDforce}a and b) it results that rescaling the energy spectrum of velocity and density result in a great scattering of \apca. When performing a linear fit by minimizing the $\chi^2$ error, it results in standard errors ranging from about 0.24 to 0.45 on velocity-modified MHD and around 0.2 on density+velocity-modified MHD simulations. This represents a big scattering when we compare with what is seen in Fig. \ref{atau-fbm} top panel for fBm simulations.

\placefigure{atau-MHDforce}

BH02a and \citet{2011ApJ...740..120R} find that \apca~has no dependence on density for different forcing of the turbulence (solenoidal and compressive) using fBm, MHD simulations or mixtures of fBm velocity with MHD density simulations. In this study we change the spectral slope of density for either fBm simulations (see Fig. \ref{abeta-fbmdn}) either modified MHD simulations (see figs. \ref{pwer-mhdforcedn} and \ref{atau-MHDforce}b).

One possible cause of this chaotic behavior is that the density also plays a very important role in PCA (see Fig. \ref{pwer-mhdcube}) and this is not possible to notice using only fBm or MHD simulations with constant density to calibrate PCA. When both velocity and density spectra are forced to have the same spectral slope, PCA should be able to relate directly to this spectral slope, but figure \ref{pwer-mhdforcedn} shows differently. We believe that the phase generated by the radiative transfer is responsible for this divergence between what we expected if PCA would relate directly to $\beta$ and this results.

\section{Discussion}\label{disc}
Using MHD simulations with radiative transfer, \citet{2013ApJ...771..123B} found that for CO integrated intensity maps the power spectrum saturates to a universal slope of $-3$, in agreement with predictions from LP04. The implication of this result is that it is impossible to recover the density power spectrum from CO in the optically thick limit. This has further implications for the use of VCS and VCA in order to find the velocity power spectrum of turbulence, particularly in the star forming molecular medium where high optical depth tracers are common. For optically thick CO, the VCA may only be successfully applied in the limit of thin velocity channels and other tracers may need to be used to find the density power spectrum.

In this paper we show that the PCA may have some advantages in the optically thick limit as it makes use of the phase information present in the data. We proved this by randomizing the phase of velocity components and observing tha PCA exponent dependences are lost (see figs. \ref{pwer-mhd}, \ref{pwer-mhdforce} and \ref{atau-fbm}). This is in contrast to the traditional power spectrum analysis that is not sensitive to the phase information. This may be an important result in the direction to initiate the development of tools that takes explicitly into account the phase information in the turbulence analysis.

The problems of PCA in achieving the goal of studying turbulence at high opacities are related to the irregular behavior of PCA exponents at high Mach numbers that we also report. Further studies of how to deal with this problem are necessary. Currently we believe that for small Mach numbers PCA can be used. Unfortunately, many molecular clouds exhibit highly supersonic turbulence. Additionally, rescaling the spectrum of density of MHD simulations produces a similar effect of disturbing PCA exponent at all opacity regimes, showing how tangled are the spectral and phase information  as well as the density and velocity in the resulting eigen images and eigen vectors.

This however does not contradict the results of \citet{2003ApJ...595..824B} and \citet{2011ApJ...740..120R}, since the first of these works used MHD simulations with randomized phases of the velocity field (which has the same effect of a fBm field with same spectral index)  and the second  of these works uses fBm velocity fields with optically thin MHD density, changing only the turbulence forcing (compressive or solenoidal) and spectral indexes of $\beta=1.9,~2.0$, in order to avoid intermittency of the velocity field affecting the study of density.  In this sense our study is different from these previous ones.

We should emphasize at this point that the mapping from position space to the velocity space that corresponds to the third axis in the PPV cubes in observations is a very non-linear process that produces a phase of its own, independently of the phases of velocity or density fields. In fact, fBm fields will have random phases and the PPV cubes after radiative transfer won't. In this case, randomizing the PPV phases is different than randomizing the phases of the velocity or density field.

For optically thin data it is advantageous to use VCA analysis which has a good theoretical foundation. At the same time as the optical density effects become important, the use of the PCA may become advantageous. For a range of optical depths the use of the thin slices within the VCA and the PCA approach is beneficial. Further research should define better the parameter space where the use of one or another technique is preferable. We feel that the same sensitivity to phase information that makes the PCA applicable to high opacity data makes the PCA index behave irregularly as the sonic Mach number increases. 

\section{Summary}\label{summary}
We performed two sets of numerical experiments in order to test the sensitivity of the PCA to opacity effects. One set of simulations employed 84 fBm simulations -- 48 with constant density and 36 with fBm density and varying spectral indexes. And the other set of numerical experiments employed 72 MHD turbulence simulations -- 12 being original density cubes; 36 with only rescaled velocity spectral index; and 24 with both density and velocity rescaled. The results of these simulations were used as the input data for the radiative transfer in order to obtain synthetic observations which we analyzed using both VCA and PCA technique. 

\begin{itemize}
\item Using MHD simulations we confirm theoretical predictions in LP04 and results of numerical study in \citet{2013ApJ...771..123B} in the sense that the spectrum of PPV maps saturates to the universal spectral slope of $-3$ in optically thick ISM and therefore does not contain the information about the underlying turbulence. As expected, randomizing the phases of these data sets did not affect the overall shape of the power spectra.
\item While at high optical depths, applying PCA to the PPV with shuffled phases from fBm and MHD simulations reveals that PCA is sensitive to both spectral and phase information of the PPV data cubes. This explains how PCA can retain information about turbulence at high optical depths while the Fourier power spectrum of the PPV data cubes does not contain this information.
\item The application of PCA to MHD data is not straightforward, however, as we report irregular behavior of the exponents that we obtain with the PCA technique when the turbulent sonic Mach number is greater than unity.
\end{itemize}

\acknowledgments 
Research activities of the Observational Astronomy Stellar Board at the Federal University of Rio Grande do Norte are supported by continuous grants from the
Brazilian agencies CNPq and FAPERN and by the INCT-INEspa\c{c}o. C.C. acknowledges a PVE/CAPES fellowship (Process $n^o88887.114342/2015-00$). B.B. acknowledges the support of the NASA Einstein Postdoctoral Fellowship. A.L. acknowledges a distinguished visitor PVE/CAPES appointment at the Physics Graduate Program at UFRN, Natal, Brazil and thanks the DFTE/UFRN for hospitality.% D.P. ackowledges.

\end{document}